\documentclass[%
reprint,
groupedaddress,
 amsmath,amssymb,
 aps,
prab,
floatfix,
]{revtex4-2}

\usepackage{graphicx}% Include figure files
\usepackage{dcolumn}% Align table columns on decimal point
\usepackage{bm}% bold math
\begin{document}

\newcommand{\I}{\mathrm{i}}
\newcommand{\E}{\mathrm{e}}
\newcommand{\D}{\,\mathrm{d}}
\newcommand{\s}{\mathrm{sign}}

\title{Landau distribution of ionization losses:\\history, importance, extensions} 
\author{Eugene Bulyak}
\email{bulyak@kipt.kharkov.ua, Eugene.Bulyak@gmail.com}

%\altaffiliation[\\Also at ]{V.N.~Karazin National University, 4 Svodody sq., Kharkiv, Ukraine}
\affiliation{National Science Center `Kharkov Institute of Physics and Technology', 1 Academichna str, Kharkiv, Ukraine\\ V.N.~Karazin National University, 4 Svodody sq., Kharkiv, Ukraine}

\author{Nikolay Shul'ga}
\email{shulga@kipt.kharkov.ua}
%\altaffiliation[Also at ]{V.N.~Karazin National University, 4 Svodody sq., Kharkiv, Ukraine}
\affiliation{National Science Center `Kharkov Institute of Physics and Technology', 1 Academichna str, Kharkiv, Ukraine\\ V.N.~Karazin National University, 4 Svodody sq., Kharkiv, Ukraine}

\date{\today}

\begin{abstract}
 The ionization losses -- the losses of energy by fast charged particles traveling through a matter -- have been under study for more than 100 years. The theoretical explanation of this process spans similar period. About 75 years ago, Lev Landau published a theoretical paper on the ionization losses, which drastically leveled up the research and still remains amongst the most cited in the field. The present note digests the history of theoretical development and attempts to clarify Landau's method of research and the function named after him.   
\end{abstract}

%\pacs{41.60.-m, 41.75.Ht}

\maketitle

\section{Introduction}
Lev Landau published  an article on ionization losses distribution \cite{landau44} more than  75 years ago. The article is one of the most (if not  the most) cited paper in the physics of charged particles interaction with media, because it conformed with the two criteria of an outstanding theoretical work:
\begin{itemize}
  \item ``Physical law should have mathematical beauty'' (P.A.M. Dirac)
  \item ``Everything should be made as simple as possible, but no simpler'' (A. Einstein)
\end{itemize}

Landau's paper considered distribution of energy losses by the ultrarelativistic charged particles due to ionization of the media. This process is characterized by a small magnitude of the energy loss per the individual act of scattering the fast electrons off the electrons of matter. The author assumed (for simplicity!) the losses being independent of the projectile energy.
This crucial assumption yielded a brilliant result: the Landau distribution density function: It is  also referred to as the straggling function, and was widely employed up to nowadays not only in the field of interaction of the fast particles with matter but also in the radiation emission in the periodic structures, etc.

In this note, we recall the essence of the Landau function emphasizing key assumptions and simplifications that allowed to construct the Landau distribution. The paper is organized as following: First, a short prehistory is discussed; then, the mathematical basis of Landau approach is presented together with the representations of the Landau straggling function. The note concludes with the extensions of the Landau method and function.  

\section{Prehistory}

All the early papers on the ionization losses  used the Thomson formula \cite{thomson10} for the Rutherford scattering \cite{rutherford11} in the Coulomb field. The spectrum of energy losses (probability density function) in such a field  is rather simple: 
$
\phi_x(\epsilon) \propto x/\epsilon^2\; ,
$
where  $\epsilon $ is the energy lost by a fast charged particle, $x$ is the thickness of the material. 

This spectrum is independent of the particle energy, but involves mathematical difficulties and restrictions of the physical model since it is  non-normalized: indefinite of the zero-moment
$
\int_0^\infty \phi_x(\epsilon)  \D \epsilon
$ 
makes all the higher moments indefinite.

\begin{figure}[htb]
 \includegraphics[width=\columnwidth]{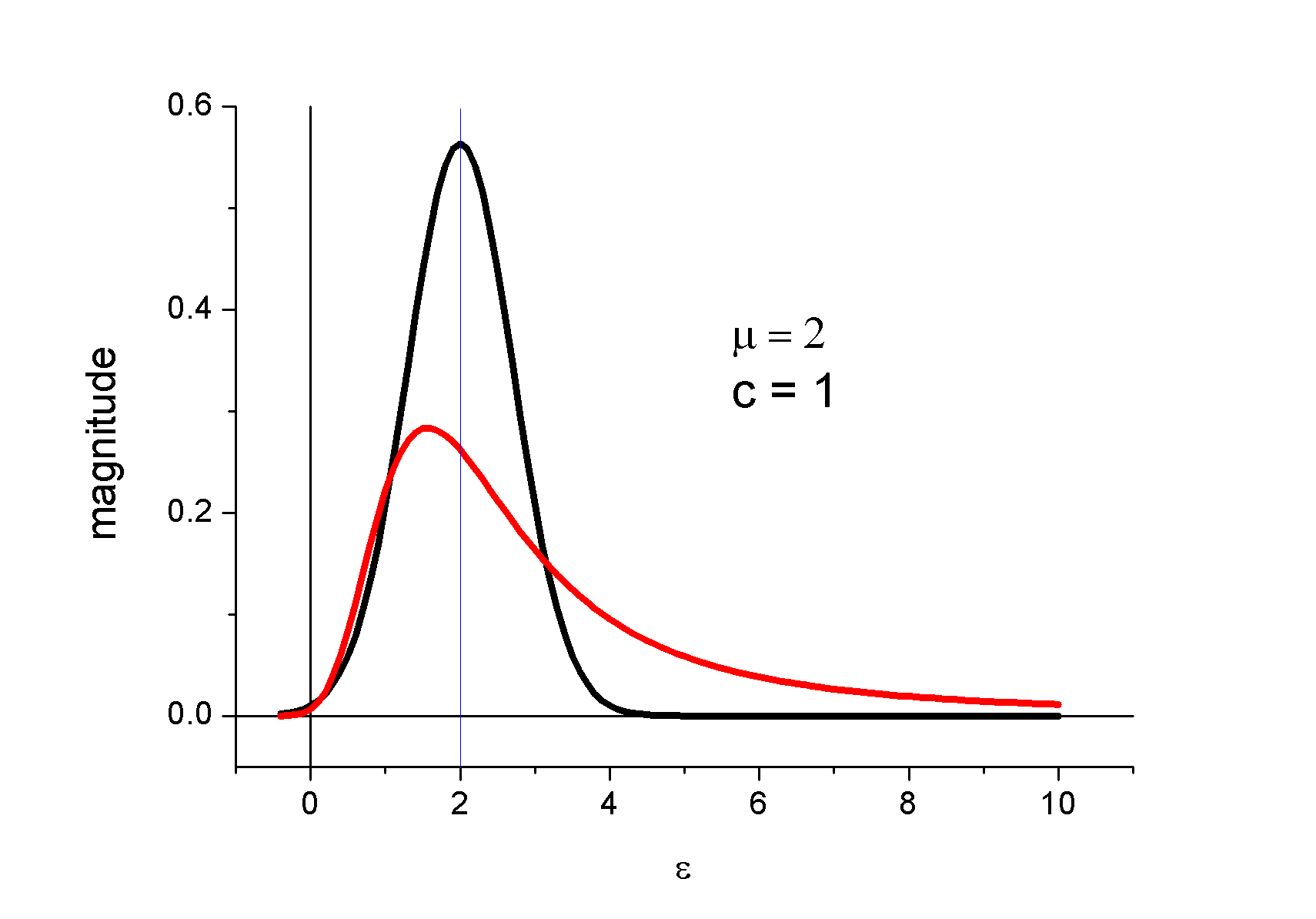}
   \caption{The Gaussian (Bohr) -- black curve and the Landau function (Williams, Landau) -- red curve.}
   \label{fig:gaulan21}
\end{figure}

All the authors who had considered the ionization losses, overcame non-normalizability in different ways. Niels Bohr \cite{bohr15} cut away the high energy tail by setting $\phi_x(\epsilon > \epsilon_\text{max}) = 0$ for a given material thickness $x$ at the energy where the number of scattering at $\epsilon = \epsilon_\text{max}$ equal  unity. He defined the minimal energy, $\epsilon_\text{min}$, such that the average energy loss  is equal to the experimental data. 
Then, he assumed the second moment being linearly proportional to the material thickness, and finally got the Gaussian distribution of the ionization losses.  Figure~\ref{fig:gaulan21} presents the Gaussian and Landau distributions for same magnitudes of the average and the scale (width of distribution). (We do not use the terms mean and variance since both are indeterminate for the Landau distribution.) The mode (maximum) of the Landau distribution is locate at smaller energy than the Gaussian, and this distribution has a heavy tail directed to higher energies.  

Evan James Williams \cite{williams29}  had criticized Bohr and indicated that Bohr's approach is applicable only for heavy particles capable of many interactions, while light particles -- electrons -- experience much fewer interactions. Williams shifted maximum energy to lower magnitudes and the remaining tail did not cut off but treated in the Poissonian way: the distribution was composed  of losses in one interaction, two interactions, and so on up to four interactions, with the Poissonian probability.

Let us present quotation from Roy and Reed \cite{roy68} on the ionization losses theories:
\begin{quotation}
The theory of the average energy loss by ionization per unit length of the electron in the stopping material has been worked out by Bethe \cite{bethe33} and Bloch, \cite{bloch33,bloch33a}, in some detail, on the basis of the electron-electron scattering cross section of M{\o}ller \cite{moller32}. It was assumed that (a) the binding energies of the atomic electrons in stopping material are much less than the kinetic energy of the incident electrons, (b) saturation of the energy loss gradient due to polarization effects in the target are negligible, and (c) the principle energy loss mechanisms are excitation and ionization, with radiative losses being negligible.

Williams \cite{williams29} followed by earlier work of Bohr \cite{bohr15} obtained an expression for the most probable energy loss of fast electrons. Later, Landau \cite{landau44} obtained, by mathematical analysis quite different from that of Williams, essentially the same results for the energy distribution.
\end{quotation}

In Grupen's paper \cite{grupen99}, started from the Landau (balance) equation and employed the Laplace transform, the author derived a Landau-like function in integral form. Also, it was pointed out that Hines  \cite{hines55}  has used a similar expansion, but was unable, using the Mellin transforms (see,e.g., \cite{whittaker96}), to obtain an exact solution \footnote{As was privately communicated by A.I. Akhiezer to one of the coauthors (NS), L.D. Landau had `invented' a Mellin-like transform for the same purpose, but not succeeded with it.}.

\section{Mathematics of the Landau straggling function}
\subsection{Kinetic equation and solution}
Mathematics start from a univariable transport equation:
\begin{align*}
  \frac{\partial f(x,\gamma)}{\partial x} = \int_{0}^{\infty} & \left[ f(x,\gamma+\epsilon)W(\gamma+\epsilon,\gamma) - \right. \nonumber \\ 
  &\left. f(x,\gamma)W(\epsilon,\gamma)\right] \D \epsilon \; ,
\end{align*}
where $f(x,\gamma) $ is the distribution (spectrum) of fast particles that traversed the medium of thickness $x$, $W(\epsilon,\gamma)$ is the probability that the particle of energy $\gamma $ lost $\epsilon $ fraction of energy over $x=1$ path length. Initial distribution is
\[
f_0(\gamma)=g(x=0,\gamma)\; ,
\]
and $\gamma>0$, $\epsilon >0 $.

\paragraph{1st assumption:} Energy losses $\epsilon$ are independent of the particle energy $\gamma $:
\[
W(\epsilon,\gamma) = w(\epsilon)\; .
\]

This assumption is a rather `physical', since the Thomson cross section  allows for a very big energy loss that can exceed the particle energy at nonzero probability.  

For deviation of energy $\Delta = \gamma_0 - \gamma $ with $\gamma_0 $ being the initial energy of the particle, the probability distribution function  $f(x,\Delta) $ will be a solution of the kinetic equation (proposed by L.D.~Landau):
\begin{equation}\label{eq:kinet}
\frac{\partial f(x,\gamma)}{\partial x} = \int_{0}^{\gamma_0} w(\epsilon) \left[ f(x,\gamma-\epsilon) - f(x,\gamma)\right] \D \epsilon \; .
\end{equation}

The Riemann integral in \eqref{eq:kinet} may be changed for the Lebesgue one, in which a \emph{known} function $w(\epsilon)=w(\epsilon ;\gamma_0)$ serves as a measure:
\begin{align}\label{eq:kinet1}
\frac{\partial f(x,\gamma)}{\partial x} &= \int_{[0,\infty)} w(\epsilon) f(x,\gamma-\epsilon) \D \epsilon  -  f(x,\gamma) C  \; ,\\
&C\equiv \int_{[0,\infty)} w(\epsilon)\D \epsilon\; .\nonumber
\end{align}

The Laplace transform %
\footnote{Actually, Landau and then Vavilov made use of a modification of the Laplace transform known as `the Laplace-Carson transform'.}
 along with the convolution theorem converts this integro-differential equation into the differential one with respect to the transform $\phi = \mathcal{L}(f)(s)$:
\begin{equation}\label{eq:difphi}
\frac{\partial \phi(x,s)}{\partial x} = \phi(x,s)\left[\mathcal{L}(w) - C\right] \; ,
\end{equation}
which has a general solution
\begin{equation}\label{eq:solphi}
\phi(x,s) = \phi(0,s)\exp\left\{ x\left[\mathcal{L}(w) - C\right]\right\} \; .
\end{equation}

The inverse transform of \eqref{eq:solphi} yields a solution for the distribution density of losses.

\paragraph{2nd assumption:} The initial distribution of losses is of the Dirac delta, $f_0(\gamma)=\delta (\gamma)$ with its Laplace transform equal to unity. The solution  becomes:
\begin{equation}\label{eq:solphi0}
f(x,\gamma ) = \mathcal{L}^{-1} \exp\left\{ x\left[\mathcal{L}(w) - C\right]\right\} \; .
\end{equation}

\paragraph{3rd key assumption:}  Since the Thomson formula for Rutherford scattering in a Coulomb field, $w(\gamma )\propto \gamma^{-2}$, that has no moments,  L.D.~Landau made use of an ingenious method: he introduced a new dimensionless variable $\lambda $,
\[
\lambda = \frac{\Delta-\Delta_\mathrm{m}}{\xi}\; ,
\]
with $\Delta_\mathrm{m}(x)$ and $\xi(x)=x w(E)$ being  the average energy loss and the width (scale) of the spectrum  in  the material of thickness $x$, respectively.
In other words, the difference between the two indeterminate variables $\Delta$ and $\Delta_\mathrm{m}$ over the third indeterminate variable $\xi $ results in the convenient dimensionless variable $\lambda $. The values for $\Delta_\mathrm{m}$ and $\xi $ are supposed to be taken from the experimental data (as N.~Bohr had done).

\subsection{Landau straggling function}
The original Landau energy-loss distribution, $f_\mathrm{L}(E,x)$, after some idealization, obtained the form:
\begin{equation}\label{eq:landaudis}
f_L(\lambda,x)=\frac{1}{\xi(x) }\frac{1}{2\pi\I}\int_{a-\I\infty}^{a+\I\infty} \E^{s\log s +\lambda s} \D s; \quad a\in \Re\; .
\end{equation}

A standard commonly accepted form of the Landau distribution is:
\begin{equation}\label{eq:fLan}
S(x;\mu,\sigma) = \frac{1}{\pi \sigma}\int_{0}^{\infty}\E^{-t}\cos\left[ t\left( \frac{x-\mu}{\sigma}\right) + \frac{2 t}{\pi}\log \left(\frac{t}{\sigma}\right)\right]\D t\; .
\end{equation}

The characteristic function (the Fourier transform) of the distribution is:
\begin{equation}\label{eq:charfu}
\phi (t;\mu,\sigma) = \exp \left( \I t \mu -\frac{2 \I \sigma t}{\pi } \log |t| - \sigma|t| \right)\; .
\end{equation}

\subsection{Properties of the Landau distribution}
The Landau distribution function (also known as the Landau straggling function) belongs to the $\alpha$--stable distribution functions proposed and studied by Paul L\'{e}vy \cite{levy25}. %(see a recent monograph \cite{nolan17}).

The characteristic function of $\alpha $-stable  process, see e.g. \cite{nolan:2017}, has a general form:
\begin{equation}\label{eq:levy}
\phi(s)=\exp\left\{ - 2\pi\I s\mu -|2\pi \sigma s|^\alpha \left[ 1-\I\beta\, \s(s)\Phi\right] \right\}\; ,
\end{equation}
with
\[
\Phi =
\begin{cases}
\tan \left(\frac{\pi \alpha}{2}\right), & \alpha\neq 1,\\
-\frac{2}{\pi}\log |s|, & \alpha = 1,
\end{cases}
\]
The parameters of the stable distribution are: $\alpha\in (0,2]$ the stability parameter,  $\sigma >0 $ the scale parameter, $\beta $  the skewness parameter, and $\mu $ the location parameter.

The Landau distribution is a particular case of this class with $\alpha = 1$ and $\beta = 1$. It means that the distribution has neither the first (mean) nor the second (variance) moments. Its scale parameter $\sigma$ -- the half-width at $1/\E $ height -- is in direct proportion to the average losses (it increases as $x^{1/\alpha}$, see \cite{bulyak19a}). The Landau distribution is normalized to unity as it follows from \eqref{eq:fLan}: the number of particles over the whole energy range is preserved.

These two statistical but nonphysical properties -- infinite first and second moments --
do not prevent the experimentalists from wide application of the Landau distribution since they use `full width at half height' and 'asymmetry' parameters, and pay less attention to the thick infinite tail.

\section{Enhancements of the Landau function}
Enhancements of the Landau function mainly involve modifications to the recoil spectrum (the Thomson formula)  fitting it to the experimental conditions.
 
\subsection{Vavilov's generalization}
P.V.~Vavilov took into  consideration  ionization energy losses of high-energy heavy particles \cite{vavilov57} and proposed modification of the Landau distribution. He presented an ``analysis and rigorous solution of the problem of ionization losses of heavy particles in `thin absorbers,' i.e., absorbers in which the ionization losses are much smaller than the initial energy of the particles.''

Now this modified distribution is referred to as the Vavilov or the Landau-Vavilov distribution.

This distribution takes into account the maximum allowed energy transfer $\epsilon_\mathrm{max}$ in a single collision of a heavy particles with an atomic electron: The probability $w(\epsilon ) $ was used
\[
x w(\epsilon ) = \frac{\xi}{\epsilon^2}\left( 1-\beta^2 \frac{\epsilon}{\epsilon_\mathrm{max} }\right)\; .
\]

The author used product of the path length in medium $x$ by the probability distribution $w(\varepsilon )$ in a form
\begin{align*}
xw(\epsilon )&= \xi \epsilon^{-2} (1-\beta^2 \epsilon/\epsilon_\text{max})\;,\\
\epsilon_\text{max}&=2 m_ec^2\beta^2/(1-\beta^2)\;,\\
\xi &= 0.300 x (m_ec^2/\beta^2)Z/A\; ,
\end{align*}
where $x$ is given in g\,cm$^{-2}$.

Now the Landau-Vavilov distribution is applied in the form, \cite{schorr73},
\begin{equation}\label{eq:vavilov}
f_\mathrm{LV}(\lambda,x)=\frac{1}{\xi(x) }\frac{1}{2\pi\I}\int_{a-\I\infty}^{a+\I\infty}\phi(s) \E^{\lambda s} \D s; \quad a\in \Re\; ,
\end{equation}
where
\begin{align*}
  \phi(s) &= \exp\left[ \kappa (1+\beta \gamma_E)\right] \exp\left[\psi(s)\right]\; ; \\
  \psi(s) &= s\log \kappa +(s+\beta^2 \kappa)\left[ \int_{0}^{1} \frac{1-\E^{-s t/\kappa} }{t}\D t - \gamma_E)\right] \\
   &- \kappa\exp ( -s/\kappa )\; ,
\end{align*}
$\kappa >0 $ is proportional to the ratio of the mean energy loss over the path length to the largest energy transfer possible in a single collision with an atomic electron, $\beta = v/c $; $\gamma_E$ is Euler's constant.%\cite{schorr73}

For a large maximum transfer energy, $ \kappa \lesssim 0.01$,  the Vavilov distribution converges to Landau's, while for a small maximum transfer, $ \kappa\gtrsim 10$ it converges to the Gaussian  \cite{seltzer64} with the mean value
\[
\mu = \gamma_E - 1 - \beta^2 - \log \kappa
\]
and variance
\[
\sigma^2 = \frac{2-\beta^2}{2 \kappa}
\]

\subsection{Bulyak-Shul'ga generalization}
The authors developed a general method for evaluating the energy spectrum evolution of relativistic charged particles that have undergone small quantum losses, such as the ionization losses, when the electrons pass through matter and the radiation losses in the periodic fields \cite{bulyak17b,bulyak18}. These processes are characterized by a small magnitude of the recoil quantum as compared with the particle's initial energy -- the recoil spectrum should have a compact support. (Actually, the requirement for a compact support is not necessary, a sufficient condition is that the recoil spectrum is square-integrable, but the compact support allows the spectrum sound physically, see \cite{bichsel88,bichsel06}.)  

As was shown in the paper \cite{bulyak17b}, under the condition of small losses,  a solution to the balance equation \cite{landau44,vavilov57} -- the characteristic function of the beam spectrum evolution -- is
\begin{equation} \label{eq:charactfun}
\hat{f}_\chi = \hat{f}_0 \E^{\chi(\check{w}-1)}\; ,
\end{equation}
where $\chi\ge 0$ is the ensemble-average number of the scattering events,  $f_0(\epsilon ) = f_{\chi=0}(\epsilon)$ is the initial spectrum,  $\hat{g}\equiv \mathcal{F}\{g\}$ denotes the Fourier transform of function $g$, and $\check{g}\equiv \mathcal{F}^{-1}\{g\}$ is the inverse Fourier transform.

A general form of the straggling function followed from \eqref{eq:charactfun} reads
\begin{equation} \label{eq:strugfun}
S_g(\epsilon ) = \E^{-\chi}\delta (\epsilon ) + \left( 1-\E^{-\chi}\right)\,S_\mathrm{BS}(\epsilon, \chi)\; ,
%\hat{f}_x = \hat{f}_0 \E^{x(\check{w}-1)}\; ,
\end{equation}
where $\delta (\epsilon )$ is the Dirac delta function,  and
\[
 \hat{S}_\mathrm{BS}(s, \chi) = \hat{w}(s)\E^{\chi\left[\hat{w}(s)-1\right]}\; .
\]

The function \eqref{eq:strugfun} adequately describes the density distribution for  the entire range of the integrated ensemble-average energy losses, from the very thin media, $\chi\ll 1 $, up to the thick $\chi \gg 1$, while the Landau-Vavilov is only valid for many recoils, $\chi \gg 1$.

\section{Conclusion}
L.D.~Landau created the method to derive the universal function named after him. The method is essentially based on two pillars: (i) advanced mathematics, and (ii) using dimensionless variables and parameters. He employed the Laplace-Carson integral transform and introduced the dimensionless variable: ratio of the lost energy deviation from the average loss to the width of the loss distribution.

The Landau function is included among other special functions in a popular \textsc{mathematica} package \footnote{\emph{https://www.wolfram.com/mathematica}}.

Both the method and the function are still widely used: the method to derive new functions, and the function to derive or validate new methods, see, e.g., \cite{ashrafi17,bulyak18,bulyak19a}.
%\bibliography{bushuEntro,bushu,bul-engl}

%apsrev4-2.bst 2019-01-14 (MD) hand-edited version of apsrev4-1.bst
%Control: key (0)
%Control: author (8) initials jnrlst
%Control: editor formatted (1) identically to author
%Control: production of article title (0) allowed
%Control: page (0) single
%Control: year (1) truncated
%Control: production of eprint (0) enabled
\providecommand{\noopsort}[1]{}\providecommand{\singleletter}[1]{#1}%\providecommand{\noopsort}[1]{}\providecommand{\singleletter}[1]{#1}%

\end{document}